\documentclass[aps,prb,twocolumn,showpacs]{revtex4}
\usepackage{amsmath}
\usepackage{graphicx}
\newcommand{\al}{\alpha}
\newcommand{\be}{\beta}
\newcommand{\g}{\gamma}
\newcommand{\de}{\delta}
\newcommand{\e}{\epsilon}

\newcommand{\thi}{\theta}

\newcommand{\ka}{\kappa}
\newcommand{\la}{\lambda}

\newcommand{\mi}{\mu}
\newcommand{\n}{\nu}
\newcommand{\p}{\pi}

\newcommand{\s}{\sigma}

\newcommand{\w}{\omega}
\newcommand{\W}{\Omega}

\newcommand{\G}{\Gamma}
\renewcommand{\S}{\Sigma}
\newcommand{\D}{\Delta}
\newcommand{\ta}{\tau}

\newcommand{\pder}[2]{\frac{\partial #1}{\partial #2}}

\newcommand{\round}[1]{\left({#1}\right)}
\newcommand{\square}[1]{\left[{#1}\right]}
\newcommand{\abs}[1]{\left|{#1}\right|}
\newcommand{\cvec}[2]{\round{\begin{array}{c}#1\\#2\end{array}}}
\newcommand{\mat}[4]{\round{\begin{array}{cc}#1&#2\\#3&#4\end{array}}}
\newcommand{\rvec}[2]{\round{\begin{array}{cc}#1&#2\end{array}}}

\newcommand{\beq}{\begin{equation}}
\newcommand{\eeq}{\end{equation}}
\newcommand{\Beq}{\begin{eqnarray}}
\newcommand{\Eeq}{\end{eqnarray}}
\newcommand{\bml}{\begin{multline}}
\newcommand{\eml}{\end{multline}}
\newcommand{\bsp}{\begin{split}}
\newcommand{\esp}{\end{split}}

\begin{document}

\title{Theory of electron-phonon interaction in a nonequilibrium open electronic system}

\author{So Takei$^1$ and Yong Baek Kim$^{1,2}$}
\affiliation{$^{1}$Department of Physics, The University of Toronto, Toronto,
Ontario M5S 1A7, Canada\\
$^2$School of Physics, Korea Institute for Advanced Study, Seoul 130-722, Korea}
\date{\today}

\begin{abstract}
We study the effects of time-independent nonequilibrium drive on an {\it open} 2D 
electron gas system coupled to 2D longitudinal acoustic phonons using the Keldysh path integral 
method. The layer electron-phonon system is defined at the two-dimensional interface 
between a pair of three-dimensional Fermi liquid leads, which act both as a particle pump 
and an infinite bath. The nonequilibrium steady state is achieved in the 
layer by assuming the leads to be
thermally equilibrated at two different chemical potentials. This subjects the layer to an out-of-plane 
voltage $V$ and drives a steady-state charge current perpendicular to the system. 
We compute the effects of small voltages ($V\ll\w_D$) on the in-plane electron-phonon scattering rate 
and the electron effective mass at zero temperature. We also find that the obtained nonequilibrium 
modification to the acoustic phonon velocity and the Thomas-Fermi screening length reveal the 
possibility of tuning these quantities with the external voltage.  
\end{abstract}
\maketitle

%%%%%%%%%%%%%%%%%%%%%%%%%%%%%%%%%%%%%%%%%%%%%%%%%%%%%%%%%%%%%%%%%%%%%%%%%%%%%%%%
\section{Introduction}
\label{introduction}
When considering steady states in a non-adiabatic {\it closed} driven system it is necessary 
for one to specify dissipation effects. Intuitively, bulk heating by an external drive must be
compensated by bulk dissipation or some form of thermal contact with an infinite bath so as
to prevent gradual heating. In the absence of a clear dissipative mechanism 
within the system one must consider some form of coupling between the system and an 
infinite reservoir through which heat can escape. However, it is difficult to model
an infinite bath that would dissipate heat at every location in a 3D bulk system. 
Recent theoretical works\cite{dalphil,feldman,greensondhi,greenetal,hogangreen} 
in electrically driven steady-state systems investigated 
universal scaling behaviour in transport quantities near various quantum critical points. 
In the work by Dalidovich {\it et al}\cite{dalphil}, a heat sink was considered by adding a 
phenomenological dissipation term in the Lagrangian as suggested by Caldeira and Leggett\cite{calleg}. 
In the work by Green {\it et al}\cite{hogangreen}, a heat sink for the itinerant electrons 
was assumed to be provided by the underlying lattice. 

In the treatment of these closed nonequilibrium systems, it is often 
the case that precise details of the coupling between such systems and their environment are not known, and
one is often reduced to describing these effects phenomenologically in an {\it ad hoc} fashion. 
An alternative method of treating the heating problem in a system is to begin with a theory which 
explicitly includes couplings between the system and its external reservoirs. In an {\it open} driven 
system these reservoirs naturally act both as a source of nonequilibrium drive (particle pump) and a 
heat sink. A vast number of theoretical works on open driven systems have been conducted in mesoscopic 
physics. Examples can be found in quantum dot systems\cite{rev,jauhoetal,rosch,mitra1} where leads that couple 
to the dot can be envisaged as the reservoirs. 

Though the role of these reservoirs is vital, it is desirable to have a final effective theory formulated
by degrees of freedom associated with the central active system. From a mathematical 
viewpoint, this requires one to trace out reservoir degrees of freedom from the starting theory. 
This process of tracing out bath degrees of freedom is most straightforwardly done in the 
language of functional integrals. In the context of nonequilibrium systems the Keldysh path 
integral method is a suitable technique\cite{keldysh,rammer,kamenev}. 
The formalism has been used previously in characterizing zero-dimensional open systems, like 
quantum dots, subject to charge currents\cite{jauhoetal,mitra1,mitra2}. In the context of {\it extended} 
open systems the formalism has recently been applied to a two-dimensional itinerant electron 
system\cite{mitra3} and a microcavity polariton system\cite{little2,little}. 

In this work we consider a steady-state open 2D metallic system which is electrically driven. 
We model the central metal layer in a jellium model in which conduction electrons
couple to 2D longitudinal acoustic phonons. Two metallic leads, which sandwich the metal layer
(Fig.\ref{fig:system}), are in thermal equilibrium at two different chemical potentials, and 
establish an out-of-plane charge current through the layer. While the electrons are driven
out of equilibrium by its direct coupling to the charge current the phonons are out of equilibrium
due to their coupling to the electrons. Using the Keldysh formalism we
investigate the effects of nonequilibrium perturbation on the properties of electrons, phonons,
and their interactions. 

We now summarize our main findings. First, we find that the zero-temperature 
phonon velocity is modified in the presence of a nonequilibrium voltage,
\beq
\de c(V)\sim(\D\G)V.
\eeq
$V=\mi_L-\mi_R$ is the nonequilibrium voltage defined as the difference between the
two chemical potentials of the leads, and $\D\G=\G_L-\G_R$ describes the asymmetry in the
coupling strengths of the central metal layer to the two leads. We note that the sound
velocity can be tuned by varying voltage $V$, and may be increased or decreased
according to the polarity of the voltage. Screening properties of electrons are also 
modified by the voltage. In particular, we note the nonequilibrium correction to the 
Thomas-Fermi wavevector, 
\beq
\de k_S\sim(\D\G)V,
\eeq
which shows how the voltage can modify the distance scale beyond which the Coulombic disturbance 
of the ions is effectively screened by the conduction electrons. We also consider modifications to
in-plane transport in the presence of out-of-plane current. In particular, the out-of-equilibrium
electron-phonon scattering rate at zero temperature, for voltages much smaller than the 
Debye frequency ($V\ll\w_D$), scales with voltage as
\beq
{1\over \ta_{el-ph}}\sim V^3.
\eeq 
We note that in thermal equilibrium, the scattering rate scales 
with respect to temperature with the same power law. The reason for
this, as discussed in the main parts of this paper, can be found in the similar responses of
the in- and out-of-equilibrium electron distribution functions to temperature and voltage,
respectively. Corrections to the electron mass enhancement factor, $\la={m^*\over m}$,
with respect to voltage at zero temperature is 
\beq
\de\la(V)\sim V^2,
\label{dela}
\eeq
with the finite-temperature correction scaling with the same exponent. Eq.\ref{dela} implies
that voltage can be used to tune the effective mass of the conduction electrons.

The paper is organized as follows. We present and describe our model in Sec.\ref{model} and
derive an effective phonon action using the Keldysh formalism in Sec.\ref{effs}. In
Sec.\ref{noneqmprop}, we compute effects of nonequilibrium perturbation on the phonon velocity,
electron-phonon scattering rate, and electron mass enhancement. We conclude in Sec.\ref{conclude}.
Details of the calculations in Sec.\ref{effs} are provided in Appendix \ref{details}.

%%%%%%%%%%%%%%%%%%%%%%%%%%%%%%%%%%%%%%%%%%%%%%%%%%%%%%%%%%%%%%%%%%%%%%%%%%%%%%%%
\section{Theory}
\subsection{Model}
\label{model}
\begin{figure}[h]
\begin{center}
\includegraphics[scale=0.8]{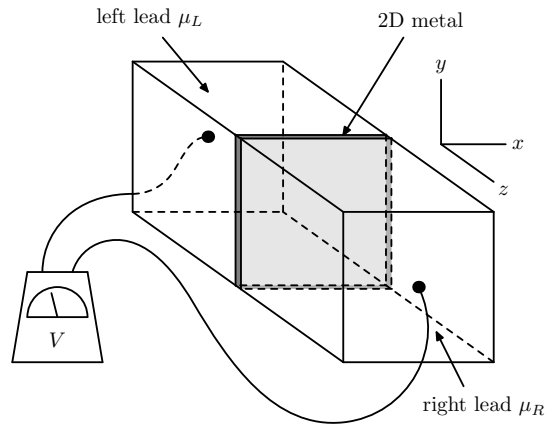}
\caption{\label{fig:system} A schematic representation of the system. A 2D
metal layer is sandwiched between two 3D leads. The leads are assumed to
be noninteracting electron gases in thermal equilibrium at temperature $T$ but may be 
held at two different chemical potentials.}
\end{center}
\end{figure}

The geometry of our system consists of two 3D electron reservoirs,
or leads, which are in contact at a two-dimensional interface. We define this 
two-dimensional interface as our central metal layer of interest (Fig.\ref{fig:system}). 
We split the Hamiltonian into three pieces: 
$H=H_L+H_t+H_{layer}$, where $H_L$  and $H_{layer}$ describe the leads and the
metal layer respectively, and $H_t$ models the tunneling between the leads 
and the layer.
We view the leads as noninteracting 3D electron gases. They are both 
in thermal equilibrium at temperature $T$ but may possess different chemical 
potentials $\mi_L$ and $\mi_R$, thus driving a steady-state current in the direction 
perpendicular to the metal layer. A continuum of states is assumed 
in the leads, occupied according to the Fermi distribution $f^0_\al(x)=\square{1+
e^{\be\round{x-\mi_\al}}}^{-1}$, where $\al\in (L,R)$ labels the leads. $H_L$ 
is then given by 
\beq
H_L=\sum_{{\bf k}_{||},k_z\atop\s,\al =L,R}\e_{{\bf k}_{||},k_z}c^{\dag}_{{\bf k}_{||},
k_z,\s,\al}c_{{\bf k}_{||},k_z,\s,\al}.
\eeq
The $z$-axis of our coordinate system coincides with the normal of the metal layer, 
which is defined on the $z=0$ plane. ${\bf k}_{||}$ denotes the component 
of a momentum vector parallel to the $z=0$ plane and $k_z$ is its component normal to
the plane. Hereafter, we drop the ``$||$'' label on all momenta for brevity, and we 
take $\hbar=1$.  

The form of our tunneling Hamiltonian $H_t$ assumes conservation of spin $\s$ and parallel
momentum vector ${\bf k}$ of the tunneling electrons, thus making $\s$ 
and ${\bf k}$ good quantum numbers throughout the 
entire lead-layer-lead structure. For computational simplicity, we assume energy-independent
tunneling matrix elements but maintain their lead-dependences in order to describe possible
asymmetries in the lead-layer couplings. These assumptions can be summarized
in the following form for the tunneling Hamiltonian,
\beq
H_t=\sum_{{\bf k},k_z\atop\s,\al =L,R}t_{\al}\round{c^{\dag}_{{\bf k},k_z,\s,\al}
d_{{\bf k},\s}+h.c.}.
\eeq

Finally, the central layer is a 2D metal whose Hamiltonian is,
\Beq
H_{layer}&=&\sum_{{\bf k},\s}\e_{{\bf k}}d^\dag_{{\bf k},\s}d_{{\bf k},\s}
+\sum_{{\bf k}}\w_{{\bf k}}\round{b^\dag_{{\bf k}}b_{{\bf k}}+\frac{1}{2}}\nonumber\\
&+&\sum_{{\bf q}\ne 0}\sum_{{\bf k},{\bf k'}\atop \s,\s '}\frac{V^{ee}_{{\bf q}}}{2}
d^\dag_{{\bf k}+{\bf q},\s}d^\dag_{{\bf k'}-{\bf q},\s '}
d_{{\bf k'},\s '}d_{{\bf k},\s}\label{hlayer}\\
&+&\sum_{{\bf k},{\bf q},\s}g_{{\bf q}}d^\dag_{{\bf k+q},\s}d_{{\bf k},\s}
u_{{\bf q}}.\nonumber
\Eeq
$V^{ee}_{{\bf q}}=2\p e^2/q$ is the 2D Coulomb potential. Our theory assumes
a jellium model for the electron-ion interaction and consequently only describes a coupling
between longitudinal acoustic phonons and electrons. The effects of transverse acoustic phonons 
are not included in this work because most important properties of a metal can be understood 
by observing the coupling of electrons to changes in the background ionic charge density which are
described by acoustic phonons. The discussion of optical phonons may be neglected given
that our work only considers monatomic lattices. In the 2D jellium model
of the electron-phonon system, the unscreened electron-phonon coupling can be shown to be
\beq
g_{{\bf q}}=-i\hat{{\bf e}}_{\bf q}\cdot{\bf q}V^{ie}_{{\bf q}}\sqrt{\frac{n_i}{2M\w_p}},
\eeq
where $\hat{{\bf e}}_{\bf q}$ is the polarization vector of the longitudinal phonons for
each ${\bf q}$ and $V^{ie}_{{\bf q}}=ZV^{ee}_{{\bf q}}$. $\w_p(q)=\sqrt{2\p n_iZ^2e^2q/M}$ is
the 2D ionic plasma frequency and  $u_{{\bf q}}=b_{{\bf q}}+b^\dag_{-{\bf q}}$
is the dimensionless phonon displacement operator. $M$, $Z$, and $n_i$ denote the mass, valence 
number, and number density per unit area of the ions respectively. 

%%%%%%%%%%%%%%%%%%%%%%%%%%%%%%%%%%%%%%%%%%%%%%%%%%%%%%%%%%%%%%%%%%%%%%%%%%%%%%
\subsection{Effective Nonequilibrium Phonon Action}
\label{effs}
In this section, we obtain a description
of the lattice vibrations in the central metal layer. The metal layer consists 
of two main components: interacting electrons and phonons. When the layer is 
coupled to two electron reservoirs with different chemical potentials, one can
think of the bath with the higher (lower) chemical potential as a particle pump (sink)
to the central active layer. Clearly, if the rate of thermalization within the 
layer is much smaller than the rate of tunneling between the layer and the leads,
the electrons in the layer ceases to follow the Fermi-Dirac distribution. 
This nonequilibrium nature of layer electrons also
influences the properties of the phonons due to electron-ion coupling. Therefore, the 
system must be described using a nonequilibrium treatment of the layer electrons, 
phonons and their interactions. We use the Keldysh formalism to provide a 
unified description of the two-dimensional electron-phonon system both in and 
out of equilibrium. 

The Keldysh formalism requires each field to possess two values at each point in time,
one on the forward branch and another on the backward branch. As a consequence, electron 
($\hat{G}$) and phonon ($\hat{D}$) Green functions, along with their respective self-energies
($\hat{\S}$,$\hat{\Pi}$), must be represented by matrices
\beq\bsp
\hat{G}&=\mat{G^R}{G^K}{0}{G^A},\hat{D}=\mat{D^K}{D^R}{D^A}{0},\\
\hat{\S}&=\mat{\S^R}{\S^K}{0}{\S^A},\hat{\Pi}=\mat{0}{\Pi^A}{\Pi^R}{\Pi^K}.
\label{keldmat}
\end{split}\eeq
For the many-particle Hamiltonian introduced in Sec.\ref{model}, the Keldysh generating
functional, $Z_K$, can be obtained by a familiar real-time coherent state functional integral
technique with the only modification arising from performing the real-time
integral over a time-loop contour\cite{kamenev,rammer}. After integrating out the lead degrees 
of freedom and introducing a Hubbard-Stratonovic field $\phi$, which decouples the two-body 
electron interaction term, the effective Keldysh generating functional yields
\beq\bsp
Z_K&=\int \mathcal{D}[d,\bar{d},u,u^*,\phi,\phi^*]e^{iS_K^{eff}[d,\bar{d},u,u^*,\phi,\phi^*]}\\
&=\int \mathcal{D}[d,\bar{d},u,u^*,\phi,\phi^*]\\
&\qquad\times e^{i\round{S^{el}_K[d,\bar{d},u,u^*,\phi,\phi^*]+S^{ch}_K[\phi,\phi^*]
+S^{ph}_K[u,u^*]}},
\end{split}\eeq
where
\beq\bsp
i&S^{el}_K[d,\bar{d},u,u^*,\phi,\phi^*]=2i\int dt\sum_{{\bf k},{\bf k'},\s}
\rvec{\bar{d}_\s^1({\bf k'},t)}{\bar{d}_\s^2({\bf k'},t)}\\
&\times\left[\hat G^{-1}({\bf k'},{\bf k},t)+\hat{\Phi}({\bf k'}-{\bf k},t)\right]
\cvec{d_\s^1({\bf k},t)}{d_\s^2({\bf k},t)},
\label{skel}
\end{split}\eeq
\beq\bsp
iS^{ch}_K[\phi,\phi^*]&=2i\int dt\sum_{{\bf k}}\rvec{\phi^{cl*}({\bf k},t)}
{\phi^{q*}({\bf k},t)}\\&\times\mat{0}{\frac{k}{2\p}}{\frac{k}{2\p}}{0}
\cvec{\phi^{cl}({\bf k},t)}{\phi^{q}({\bf k},t)},
\end{split}\eeq
\beq\bsp
i&S^{ph}_K[u,u^*]=2i\int dt\sum_{{\bf k}}\rvec{u^{cl*}({\bf k},t)}{u^{q*}({\bf k},t)}\\
&\times \hat{D}_0^{-1}({\bf k},t)\cvec{u^{cl}({\bf k},t)}{u^{q}({\bf k},t)}.
\end{split}\eeq
In the absence of leads, hence for a closed system in equilibrium, the inverse 
Green function matrix, 
\beq
\hat G^{-1}({\bf k'},{\bf k},t)=\mat{G_R^{-1}({\bf k},t)\de_{{\bf k'}{\bf k}}}
{[\hat{G}^{-1}({\bf k},t)]^{K}\de_{{\bf k'}{\bf k}}}{0}{G_A^{-1}({\bf k},t)\de_{{\bf k'}{\bf k}}},
\eeq
describing the propagation of electrons in the central layer has the 
usual free-fermion form. However, these Green functions are modified due to tunneling 
between the layer and the leads. For energy-independent tunneling
amplitudes $t_{L,R}$, the dressed Fermion Green functions were calculated elsewhere\cite{mitra1,mitra2},
\beq\bsp
G^R(K)&=\frac{1}{\w-\e_{\bf k}+i\G}=[G^{A}(K)]^*\\
G^K(K)&=\frac{-2i\sum_\al\G_\al\tanh\round{\frac{\w-\mi_\al}
{2T}}}{(\w-\e_{\bf k})^2+\G^2}.
\end{split}\eeq
Here, $\G=\G_L+\G_R$, and $\G_\al=\p\n t_\al^2$ is the rate at which electrons escape from 
the active layer into lead $\al$. $\n=m/\p$ is the 2D density of states (including both spin
projections) for electrons at Fermi energy. We have also introduced an abbreviation, $K=({\bf k},\w)$.
The $\hat{\Phi}$ matrix in Eq.\ref{skel} encodes the coupling of layer electrons to
both the Hubbard-Stratonovic field and phonons. It is given by
\beq
\hat{\Phi}({\bf k},t)=\mat{\Phi^{cl}({\bf k},t)}{\Phi^{q}({\bf k},t)}
{\Phi^{q}({\bf k},t)}{\Phi^{cl}({\bf k},t)},
\eeq
where
\beq\bsp
\Phi^{cl}({\bf k},t)&=e\phi^{cl}({\bf k},t)-g_{{\bf k}}u^{cl}({\bf k},t),\\
\Phi^{q}({\bf k},t)&=e\phi^{q}({\bf k},t)-g_{{\bf k}}u^{q}({\bf k},t).
\label{Phi}
\end{split}\eeq
The inverse phonon Green function matrix, in the absence of electron-ion coupling, is given 
by
\beq
\hat{D}_0^{-1}({\bf k},t)=\mat{0}{[D_0^A({\bf k},t)]^{-1}}{\square{D_0^R({\bf k},t)}^{-1}}
{\square{\hat{D}_0^{-1}({\bf k},t)}^{K}}.
\eeq
$\hat{D}_0({\bf k},t)$ describes the collective modes of the lattice ions as they oscillate
in a negatively charged static background of electrons at the 2D ionic plasma frequency, 
$\w_p(q)$. The bare retarded and Keldysh propagators have the usual form,
\beq\bsp
D^R_0(Q)&=\frac{2\w_p(q)}{\W-\w_p(q)+isgn(\W)\de}=[D^A_0(Q)]^*\\
D^K_0(Q)&=-2\p i\coth\round{\W\over 2T}\\
&\times \square{\de(\W+\w_p(q))+\de(\W-\w_p(q))},
\end{split}\eeq
where $Q=({\bf q},\W)$ is a phonon momentum-energy 3-vector.

\begin{figure}[h]
\begin{center}
\includegraphics[scale=0.6]{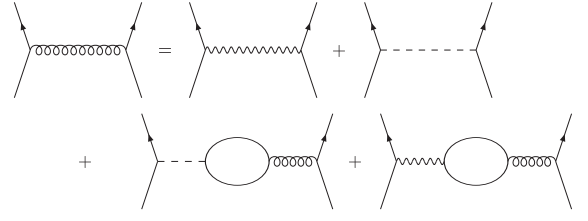}
\caption{\label{fig:coul} A diagrammatic representation of the propagator $D_\Phi$. 
The propagator, which corresponds to the curly line in the figure, describes the
effective interaction between two electrons.}
\end{center}
\end{figure}

The next step is to integrate out the layer electron fields. We use the 
standard RPA treatment for both Coulomb and electron-phonon interactions. 
A schematic representation of the resulting action is
\bml
iS^{eff}_K[u_{cl},u_q,\phi_{cl},\phi_q]\\=2i\int\frac{d^2{\bf q}}{(2\p)^2}\int\frac{d\W}{2\p}
\vec{\Phi}^{T*}(Q)\hat{D}_\Phi^{-1}(Q)\vec{\Phi}(Q),
\label{suphi}
\end{multline}
where  
\beq
\vec{\Phi}^T(Q)=\rvec{\Phi_{cl}(Q)}{\Phi_q(Q)}.
\eeq
The detailed form of this action can be found in the Appendix \ref{details}.  
The propagator matrix $\hat{D}_\Phi(Q)$ represents the effective
electron-electron interaction. Two electrons can be scattered by the unscreened
electron-electron interaction (wavy line) or by sending a phonon from one to the other
(dashed line) (see Fig.\ref{fig:coul}). In the
RPA approximation, we consider diagrams where one or more polarization bubbles are 
inserted in a chain. Both the interaction to the right and left of each of the bubbles may 
either be a Coulomb line or a phonon line. All of these diagrams are summed exactly in the
path integral formalism by solving for the inverse of the propagator.  
Within the RPA, the effective phonon action can be obtained
exactly by integrating out the Hubbard-Stratonovic field $\phi$:
\bml
iS_K^{eff}[u_{cl},u_q]\\=2i\int\frac{d^2{\bf q}}{(2\p)^2}\int\frac{d\W}{2\p}
\vec{u}^{T*}(Q)\hat{D}^{-1}(Q)\vec{u}(Q),
\label{su}
\end{multline}
where  
\beq
\vec{u}^T(Q)=\rvec{u_{cl}(Q)}{u_q(Q)}.
\eeq
As shown in the Appendix \ref{details}, the inverse phonon propagator matrix $\hat{D}^{-1}$ possesses 
the bosonic causality structure in Keldysh space. Its components are given
by
\beq
D_R^{-1}(Q,V)=2\square{D_{0R}^{-1}(Q)-\frac{|g_{{\bf q}}|^2
\Pi_R(Q,V)}{\e(Q,V)}},\label{dr}
\eeq
\bml
[\hat{D}^{-1}(Q,V)]^K\\
=2\square{[\hat{D}_0^{-1}(Q)]^K-\frac{\abs{g_{\bf q}}^2\Pi^K(Q,V)}{\e(Q,V)}}.
\label{dk}
\end{multline}
where the nonequilibrium polarization functions are given by
\bml
\Pi^R(K,V)=\frac{-i}{2}\int_Q\left[G^K(K+Q)G^A(Q)\right.\\
\left.+G^R(K+Q)G^K(Q)\right]=\Pi^{A*}(K,V)\label{pir}
\end{multline}
\bml
\Pi^K(K,V)=\frac{-i}{2}\int_Q\left[G^K(K+Q)G^K(Q)\right.\\
\left.+G^R(K+Q)G^A(Q)+G^A(K+Q)G^R(Q)\right].\label{pik}
\end{multline}
Here, $\int_Q=\int d^2{\bf q}d\W/(2\p)^3$. Voltage $V$ is defined by $V=\mi_L-\mi_R$. 
$\e(Q,V)$ denotes the nonequilibrium dielectric function which is defined by
\beq
\e(Q,V)=1-V^{ee}_{{\bf q}}\Pi_R(Q,V).
\label{dielectric}
\eeq
We see in Eqs.\ref{dr},\ref{dk} that the voltage dependence of nonequilibrium phonon 
propagators is encoded in the self-energy part; this is consistent with the fact that
phonons are driven out of equilibrium only through its coupling to the electrons. 
The Keldysh phonon propagators (Eqs.\ref{dr},\ref{dk}) and the nonequilibrium polarization 
bubbles (Eqs.\ref{pir},\ref{pik}) will now be used to calculate various properties of our 
nonequilibrium electron-phonon system. In particular,
we compute the effects of voltage on the phonon velocity, electron-phonon scattering
rate and electron mass enhancement, and compare the results with the corresponding results 
in thermal equilibrium.

%%%%%%%%%%%%%%%%%%%%%%%%%%%%%%%%%%%%%%%%%%%%%%%%%%%%%%%%%%%%%%%%%%%%%%%%%%%%%%%%%%
\section{Properties of the Nonequilibrium Electron-Phonon System}
\label{noneqmprop}
\subsection{Preliminary Considerations}

There are four energy scales in our system: temperature/voltage $T/V$,
the Debye frequency $\w_D$, the layer-lead scattering rate $\G$ and the average
Fermi energy of the two leads $\mi$. Clearly, the mass difference between ions 
and electrons lead to a large discrepancy between the dynamic time scales of these species,  
i.e. $\w_D\ll\mi$. Due to this difference, we are justified to employ the 
adiabatic approximation for the electrons in which we assume that 
electrons take on the configurations they would have if the ions were frozen into
their instantaneous positions. 
In this approximation, a frequency-independent (static) dielectric function can be used.

We assume that the layer-lead scattering rate satisfies $\w_D\ll\G\ll\mi$. 
Working in this limit validates an expansion with respect to parameter, $\W/\G\ll 1$,
where $\W$ is the typical energy scale of the excitations. The limit is consistent with the 
necessity of a strong layer-lead coupling in order for electrons to remain 
out of equilibrium midst other thermalizing effects within the layer. 
In all of our calculations, we assume $\mbox{max}\{T,V\}$ to be small 
compared to all other energy scales, 
or more precisely, $\mbox{max}\{T,V\}\ll\w_D$. Another observation one can make is 
our use of a jellium model for the ions. This sets a natural cutoff in momentum at
$1/a\sim q_F$, where $a$ is the lattice constant, because any details corresponding
to length scale $a$ or less is washed away by this continuum treatment of the lattice.

In the long-wavelength ($q\ll k_F$), small-frequency ($\W\ll\G$) limit, 
and for small voltages ($V\ll\G$), the polarization functions Eqs.\ref{pir},\ref{pik}
are given by
\beq
\Pi^R({\bf q},\W)\approx -c_1(V)+c_2(V)\round{\frac{q}{k_F}}^2-ic_3(V)\frac{\W}{\G},
\label{pir2}
\eeq
\bml
\Pi^K({\bf q},\W)\approx -ic_4\sum_{\al,\be}\frac{\G_\al\G_\be}{\G^2}\\
\times\coth\round{\frac{\W+\mi_\al-\mi_\be}{2T}}\frac{\W+\mi_\al-\mi_\be}{\G}.
\label{pik2}
\end{multline}
Here, $\mi\equiv\round{\mi_L+\mi_R}/2$ is the average chemical potential of the 
two leads. All coefficients ($c_1(V)$, $c_2(V)$, $c_3(V)$, and $c_4$) are positive, 
real, and satisfies $c_1(V)\sim c_3(V)\sim c_4\sim\n\gg c_2(V)$. $D(\mi)$ is the 
density of states at $\mi$ in the presence of leads, and is given by
\beq
D(\mi)=\frac{\n}{\p}\square{\frac{\p}{2}+\tan^{-1}\round{\frac{\mi}{\G}}}.
\label{dos}
\eeq

%%%%%%%%%%%%%%%%%%%%%%%%%%%%%%%%%%%%%%%%%%%%%%%%%%%%%%%%%%%%%%%%%%%%%%%%%%%%%%%%%%%%%%
\subsection{Nonequilibrium Bohm-Staver Relation}
An explicit expression for the dressed retarded phonon propagator calculated in Eq.\ref{dr} is
\beq
D_R(Q)=\frac{\w_p}{\W^2-\w_p^2-\frac{2\w_p|g_{{\bf q}}|^2
\Pi_R(Q,V)}{\e(Q,V)}},
\eeq
which in the long-wavelength, static limit is reduced to
\beq
D_R(Q)=\frac{\w_p}{\W^2+\frac{q\w_p^2}{2\p e^2\Pi_R(Q=0,V)}}.
\eeq
The pole of this propagator gives the dispersion of the non-attenuating longitudinal
acoustic phonon mode,
\beq
\W_{{\bf q}}^2=c^2(V)q^2,
\label{phdisp}
\eeq
where the voltage-dependent phonon velocity is,
\beq
c^2(V)=-\frac{Z^2n_i}{M\Pi_R({\bf 0},0,V)}.
\label{neqphvel}
\eeq
At zero temperature, the static, long-wavelength polarization function 
as a function of voltage is
\bml
\Pi_R({\bf 0},0,V)=-\sum_{\al}\frac{\G_\al}{\G}D(\mi_\al)\\
=-D(\mi)\square{1-\frac{\n}{2\p D(\mi)}\frac{1}{1+{\mi^2\over\G^2}}
\frac{\D\G}{\G}{V\over\G}+O\round{{V^2\over\G^2}}},
\label{pi}
\end{multline}
where $\D\G=\G_L-\G_R$. From Eqs.\ref{neqphvel},\ref{pi}, the nonequilibrium phonon
velocity is given by
\beq
c^2(V)\approx c_0^2\square{1+\frac{\n}{2\p D(\mi)}\frac{1}{1+{\mi^2\over\G^2}}
\frac{\D\G}{\G}{V\over\G}},
\label{neqphvel2}
\eeq
where
\beq
c_0^2\equiv\frac{Z^2n_i}{MD(\mi)}
\label{phvel}
\eeq
is the equilibrium phonon velocity as obtained by Bohm and Staver\cite{bohmstaver}. 
We notice in Eq.\ref{neqphvel2} that the acoustic phonon velocity can be tuned by
voltage. In particular, depending on the tunneling asymmetry and polarity, the change in
the velocity with respect to its equilibrium value may be positive or negative.

The equilibrium phonon velocity Eq.\ref{phvel} may not be in the most familiar form. 
A more common form of the equilibrium phonon velocity is derived from requiring
charge neutrality $n_e=Zn_i$ and the relationship of ratio $n_e/D(\mi)$ to the Fermi energy. 
In the considered nonequilibrium system such relations do not 
hold due to charge accumulation and an absence of a well-defined Fermi energy in the metal 
layer. However, Eqs.\ref{neqphvel},\ref{phvel} suggest that both equilibrium
and nonequilibrium phonon velocities have the form of Eq.\ref{phvel} where 
nonequilibrium effects are encoded in the generalized density of states, $-\Pi_R({\bf 0},0,V)$. 
In equilibrium, the density of states is evaluated at one single chemical potential 
$\mi$ which is located at the Fermi distribution edge. At finite voltage, the electron 
distribution function can be shown to have a split-step shape
\beq
f^{neq}(\w)=xf^0(\w-\mi_L)+(1-x)f^0(\w-\mi_R),
\eeq
\begin{figure}[h]
\begin{center}
\includegraphics[scale=0.8]{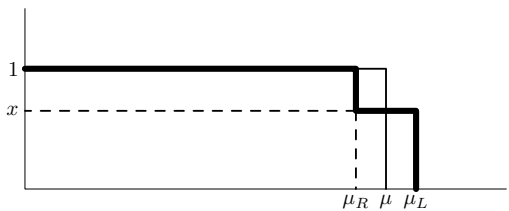}
\caption{\label{fig:nfs} Nonequilibrium steady-state electron distribution function for our system with
$\mi_L-\mi_R=V>0$ and $x=\G_L/\G$ is shown above in bold (in the absence of electron-phonon interaction
effects). The Dirac-Fermi distribution function is represented with a thin line.}
\end{center}
\end{figure}
where $f^0$ denotes the Dirac-Fermi distribution and $x=\G_L/\G$. The nonequilibrium
analog of the density of states at Fermi energy is the sum of density of states at $\mi_L$ 
and $\mi_R$ weighted by the respective lead-layer coupling strengths, and is given by:
\beq
-\Pi_R({\bf 0},0,V)=xD(\mi_L)+(1-x)D(\mi_R).
\eeq 
This voltage dependence in $-\Pi_R({\bf 0},0,V)$ is the origin of the voltage dependence seen 
in the nonequilibrium phonon velocity.

%%%%%%%%%%%%%%%%%%%%%%%%%%%%%%%%%%%%%%%%%%%%%%%%%%%%%%%%%%%%%%%%%%%%%%%%%%%%
\subsection{DC Electron-Phonon Scattering Rate}
Here, we discuss the electron scattering rate due to longitudinal acoustic
phonons. In equilibrium, at temperatures well below the Debye temperature, phonon wavevectors
are of order $T/c_0$ or less. Then, within the wavevector space of phonons that are permitted to be
absorbed or emitted by conservation laws, only a subspace of linear dimensions 
proportional to $T$ can actually participate. In two spatial dimensions, this subspace 
is one dimensional
so the phase space of allowed wavevectors is proportional to $T$. Combining this with the
fact that the screened electron-phonon coupling $|g_{sc}|^2\sim q\sim T$, we expect the
electron-phonon scattering rate to decline as $1/\ta_{el-ph}\sim T^2$. This is what is 
expected in a closed system. In the considered open system, electrons can scatter into one of the 
leads frequently so that the chance of an electron encountering a phonon and scattering
with it is decreased. Therefore, we expect a further decline in the electron-phonon
scattering rate. We then extend our calculations to the zero-temperature
finite voltage case where the scaling behaviour of the scattering rate with respect to voltage 
is determined. It is crucial to note that the behaviour of phonons depends most 
heavily on the nonequilibrium distribution function of the electrons. 
In this respect, the $T=0$, $V\ne 0$ situation 
is similar to the $T\ne 0$, $V=0$ situation in that the latter involves a smearing
of the step distribution while the former results in a split-step distribution as shown in
Fig.\ref{fig:nfs}. Therefore, phonons are subject to electrons whose distribution
behaves similarly in both situations, and one may intuitively predict
the scaling behaviours of the electron-phonon scattering rate with respect to $T$ and $V$ 
to be similar. We now quantify this possible similarity.

\begin{figure}[h]
\begin{center}
\includegraphics[scale=0.8]{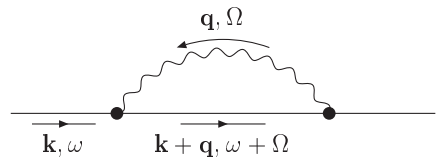}
\caption{\label{fig:SE} One-phonon self energy of the electron.}
\end{center}
\end{figure}

The starting point of this calculation is the one-loop self energy of the electron 
(Fig.\ref{fig:SE}). 
We may invoke the Migdal theorem and ignore vertex corrections. Formally, the 
self-energy in Fig.\ref{fig:SE} is given by
\bml
\S^R(K)=2i\int_Q\frac{\abs{g({\bf q})}^2}{\abs{\e(Q,V)}^2}\left(G^R(K+Q)D^K(Q)\right.\\
\left.+G^K(K+Q)D^R(Q)\right).
\label{sigma}
\end{multline}
In the static, long-wavelength ($q\ll k_F$) limit the dielectric function is 
purely real and is given by
\beq
\e_q(V)=\frac{2\p e^2c_1(V)}{q}\equiv\frac{k_S(V)}{q},
\eeq
where $k_S(V)\equiv 2\p e^2c_1(V)$ is the 2D nonequilibrium analog of the Thomas-Fermi 
wavevector. At $V=0$, $c_1$ reduces to $D(\mi)$ and the 2D equilibrium expression for 
the Thomas-Fermi wavevector, $k_S^0=2\p e^2D(\mi)$, is recovered. We notice that the voltage 
dependence observed in the Thomas-Fermi wavevector indicates that voltage can control the 
effectiveness of conduction electrons in screening Coulomb interactions between ions. Using 
$\Pi^K$ in the static approximation,
\bml
\Pi^K(Q,V)\\\approx -ic_4\sum_{\al,\be}\frac{\G_\al\G_\be}{\G^2}\coth\round{
\frac{\W+\mi_\al-\mi_\be}{2T}}\frac{\mi_\al-\mi_\be}{\G},
\label{piksl}
\end{multline} 
the dressed phonon Green functions in the static limit can be computed as follows:
\beq
D^R(Q)={\w_p \over \w_q(V)}\frac{\w_q(V)}{\W^2-\w_q(V)^2+isgn(\W)\de}
\label{dr2}
\eeq
and
\begin{widetext}
\Beq
D^K(Q)&=&-\frac{[D^{-1}(Q)]^K}{[D^R(Q)]^{-1}[D^A(Q)]^{-1}}\nonumber\\
&=&2\left[2i\coth\round{\frac{\W}{2T}}Im{[D_0^R(Q)]^{-1}}-\frac{\abs{g_{\bf q}}^2\Pi^K(Q,V)}
{\abs{\e(Q,V)}^2}\frac{ImD^R(Q)}{Im[D^R(Q)]^{-1}}\right]\nonumber\\
&=&-\p i\coth\round{\frac{\W}{2T}}\frac{\w_p}{\w_q(V)}
\round{\de(\W-\w_q(V))-\de(\W+\w_q(V))}\nonumber\\
&\qquad&\qquad\qquad+\frac{\w_p}{\w_q(V)}\frac{-2ic_4\w_p\w_q(V)\abs{g_{\bf q}}^2\sum_{\al,\be}
\frac{\G_\al\G_\be}{\G^2}\coth\round{\frac{\W+\mi_\al-\mi_\be}{2T}}\frac{\mi_\al-\mi_\be}
{\G}}{\e_q^2(V)\abs{\W^2-\w_q(V)^2+isgn(\W)\de}^2}.
\label{dk2}
\Eeq
\end{widetext}
In equilibrium, $\Pi^K=0$. In this case, by comparing the first term of Eq.\ref{dk2} 
with Eq.\ref{dr2}, we can easily verify that the fluctuation dissipation theorem is 
satisfied. 

Now one can use Eq.\ref{sigma} and the relevant Green functions to calculate the 
general DC scattering rate valid both in and out of equilibrium. Defining the static 
screened electron-phonon coupling constant as
\beq
\abs{g_{sc}({\bf q},V)}^2\equiv\frac{\abs{g_{\bf q}}^2\w_p}{\e_q^2(V)\w_q(V)},
\eeq
we obtain
\bml
\S^R({\bf k}={\bf k_F},\w=\mi,T,V)=\S^R(T,V)\\
=\S^R_1(T,V)+\S^R_2(T,V),
\label{sigmar}
\end{multline}
where
\begin{widetext}
\Beq
\S_1^R(T,V)&=&2i\int\frac{d^2{\bf q}}{(2\p)^2}\frac{d\W}{2\p}\abs{g_{sc}({\bf q},V)}^2
\frac{1}{\W-v_Fq\cos\thi+i\G}\left[-\p i\coth\round{\frac{\W}{2T}}
\round{\de(\W-\w_q(V))-\de(\W+\w_q(V))}\right.\nonumber\\
&+&\left.\frac{-2ic_4\w_p\w_q(V)\abs{g_{\bf q}}^2\sum_{\al,\be}
\frac{\G_\al\G_\be}{\G^2}\coth\round{\frac{\W+\mi_\al-\mi_\be}{2T}}\frac{\mi_\al-\mi_\be}
{\G}}{\e_q^2(V)\abs{\W^2-\w_q(V)^2+isgn(\W)\de}^2}\right],
\label{s1}
\Eeq
and
\beq
\S_2^R(T,V)=2i\int\frac{d^2{\bf q}}{(2\p)^2}\frac{d\W}{2\p}\abs{g_{sc}({\bf q},V)}^2
\frac{-2i\sum_\al\G_\al\tanh\round{\frac{\W-V_\al/2}{2T}}}
{\round{\W-v_Fq\cos\thi}^2+\G^2}\frac{\w_q(V)}{\W^2-\w_q(V)^2+isgn(\W)\de}.
\label{s2}
\eeq
Here, we have linearized the Fermion dispersion and $V_{L,R}=\pm V$ where $V=\mi_L-\mi_R\ge 0$
is assumed. Eqs.\ref{s1},\ref{s2} are the central results in the remaining sections of 
our work. In equilibrium (i.e. $\mi_L=\mi_R$), Eqs.\ref{s1},\ref{s2} reduce to
\beq
\S_1^R(T)=\int\frac{qdqd\thi}{(2\p)^2}d\W\abs{g_{sc}({\bf q})}^2
\frac{\coth\round{\frac{\W}{2T}}\square{\de(\W-\w_q)-\de(\W+\w_q)}}
{\round{\W-v_Fq\cos\thi+i\G}},
\label{sth1}
\eeq  
and
\beq
\S_2^R(T)=4\G\int\frac{qdqd\thi}{(2\p)^2}\frac{d\W}{2\p}\abs{g_{sc}({\bf q})}^2
\frac{\tanh\round{\frac{\W}{2T}}}{\round{\W-v_Fq\cos\thi}^2+\G^2}
\frac{\w_q}{\W^2-\w_q^2+isgn(\W)\de}.
\label{sth2}
\eeq
In the presence of voltage (at zero temperature),
\Beq
\S_1^R(V)&=&2\int\frac{d^2{\bf q}}{(2\p)^2}\frac{d\W}{2\p}\abs{g_{sc}({\bf q},V)}^2
\frac{1}{\W-v_Fq\cos\thi+i\G}\left[\p sgn(\W)\round{\de(\W-\w_q(V))-\de(\W+\w_q(V))}
\right.\nonumber\\
&+&\left.\frac{2c_4\w_p\w_q(V)\abs{g_{\bf q}}^2\frac{\G_L\G_RV}{\G^3}
\round{sgn(\W+V)-sgn(\W-V)}}{\e_q^2(V)\abs{\W^2-\w_q(V)^2+isgn(\W)\de}^2}\right],
\label{se1}
\Eeq
and
\beq
\S_2^R(V)=4\int\frac{d^2{\bf q}}{(2\p)^2}\frac{d\W}{2\p}\abs{g_{sc}({\bf q},V)}^2
\frac{\sum_\al\G_\al sgn\round{\W-V_\al/2}}{\round{\W-v_Fq\cos\thi}^2+\G^2}
\frac{\w_q(V)}{\W^2-\w_q(V)^2+isgn(\W)\de}.
\label{se2}
\eeq
\end{widetext}

%%%%%%%%%%%%%%%%%%%%%%%%%%%%%%%%%%%%%%%%%%%%%%%%%%%%%%%%%%%%%%%%%%%%%%%%%%%%%%%%%%%%
\subsubsection{Thermal Equilibrium: $T\ne 0$, $V=0$}
In equilibrium scattering rate calculations, temperature $T$ defines a natural cutoff 
for the momentum integrals. This is because only phonons of frequencies comparable or less
than $T$ can be absorbed or emitted. For $T\ll\w_D$, the phonon dispersion obeys 
$\w_q=c_0q$ which implies that phonon wavevectors are of order $T/c_0$ or less. We
now proceed with the integrals in Eqs.\ref{sth1},\ref{sth2}. The angular integration can
be done straightforwardly. For $\frac{\w+\W}{\G}\ll 1$,
\bml
\int_0^{2\p}\frac{d\thi}{2\p}\frac{1}{\w+\W-v_Fq\cos\thi+i\G}\\
\approx -if_0(q)+f_1(q)\frac{\w+\W}{\G},
\label{angint}
\end{multline}
where
\beq
f_0(q)=\frac{1}{\G}\frac{1}{\square{1+\round{\frac{v_Fq}{\G}}^2}^{1/2}},
\eeq
\beq
f_1(q)=\frac{1}{\G}\frac{1}{\square{1+\round{\frac{v_Fq}{\G}}^2}^{3/2}}.
\label{f1}
\eeq
Using this result one obtains,
\Beq
-Im\S^R_1(T)
&\approx&\frac{2}{\G}\int_0^{T/c_0}\frac{qdq}{2\p}\abs{g_{sc}({\bf q})}^2
\coth\round{\frac{c_0q}{2T}}\nonumber\\
&=&\frac{GT^3}{\G c_0^3}\underbrace{2\int_0^1\frac{y^2dy}{2\p}\coth\round{\frac{y}{2}}}
_{\g^{eq}_1}\nonumber\\
&=&\frac{G\g^{eq}_1}{\G c_0^3}T^3,
\Eeq
where $\g^{eq}_1\sim O(1)$ and
\beq
G(V)=\frac{\abs{g_{sc}({\bf q},V)}^2}{q}
\eeq
is a real quantity independent of $Q$. We have also used the fact that in the region of $q$ 
over which the integral is conducted, the dielectric function and $f_0$ can be approximated by
\beq
\e(q)\approx\frac{k_S^0}{q}\qquad\qquad f_0(q)\approx \frac{1}{\G}.
\eeq
Similarly, we get 
\Beq
-Im\S^R_2(T)
&\approx&\frac{1}{\G}\int_0^{T/c_0}\frac{qdq}{2\p}\abs{g_{sc}({\bf q})}^2
\tanh\round{\frac{c_0q}{2T}}\nonumber\\
&=&\frac{GT^3}{\G c_0^3}\underbrace{\int_0^{1}\frac{y^2dy}{2\p}\tanh\round{\frac{y}{2}}}
_{\g^{eq}_2}\nonumber\\
&=&\frac{G\g^{eq}_2}{\G c_0^3}T^3,
\Eeq
where $\g^{eq}_2\sim O(1)$. Therefore, in thermal equilibrium,
\beq
{1\over\ta_{el-ph}^{eq}}=\frac{G\g^{eq}}{\G c_0^3}T^3,
\label{tesc}
\eeq
where $\g^{eq}\sim O(1)$.

%%%%%%%%%%%%%%%%%%%%%%%%%%%%%%%%%%%%%%%%%%%%%%%%%%%%%%%%%%%%%%%%%%%%%%%%%%%%%
\subsubsection{Out of Equilibrium: $V\ne 0$, $T=0$}
\label{neqmsr}
In the nonequilibrium case, the momentum cutoff is set by the voltage, 
i.e. $q_c\sim V/c(V)$. Therefore, the DC scattering rate is at least of order $V$, 
and thus, we may 
set $V=0$ for the quantities in the integrand such as the phonon velocity and
dielectric function. This should allow us to obtain the scattering rate to leading 
order in $V$. When the first integral Eq.\ref{se1} is na\"{i}vely evaluated we find that the
second term, which becomes finite out of equilibrium, gives a divergent contribution
to the scattering rate. This divergence can be cured by dropping the 
static limit assumption and expanding the dielectric function to lowest order in 
frequency,
\beq
\e(Q)\approx 1+c_1\frac{2\p e^2}{q}+ic_3\frac{2\p e^2}{q}\frac{\W}{\G}.
\eeq
Recall that for $\mi\gg\G$, 
\beq
c_1\approx c_4\approx 2c_3\approx\n.
\eeq
Therefore, we get
\beq
\e(Q)\approx\frac{k_S^0}{q}\round{1+i\frac{\W}{2\G}},
\label{dfns}
\eeq
where $k_S^0\equiv 2\p e^2\n$. This new expression for the dielectric function assumes
that typical excitation energy scale is smaller than $\G$. The phonon dispersion then
becomes dissipative:
\beq
\tilde{\w}_q^2=\frac{\w_p^2(q)}{\e(Q,V)}\approx\w_q^2\round{1-i\frac{\W}{2\G}},
\eeq 
where $\w_q$ is phonon dispersion in the static limit. The modified phonon Green functions 
are,
\beq
D^R(Q)=\frac{\w_p}{\W^2-\w_q^2(V)+i\w_q^2(V)\frac{\W}{2\G}},
\eeq
\beq
D^K(Q)=\frac{2\abs{g_{\bf q}}^2\Pi^K(Q,V)}{\abs{\e_q(\W,V)}^2
\w_q^4(V)}\frac{\w_p^2}{\round{1-\frac{\W^2}{\w_q^2(V)}}^2+\frac{\W^2}{4\G^2}}.
\eeq
Now we are set to compute the first self-energy term,
\begin{widetext}
\beq
-Im\S_{1}^R(V)=4\int\frac{qdq}{2\p}\frac{d\W}{2\p}\abs{g_{sc}({\bf q})}^4f_0(q)
\frac{-Im\Pi^K(Q)}{\w_q^2\round{1+\frac{\W^2}{4\G^2}}^2}
\frac{1}{\round{1-\frac{\W^2}{\w_q^2}}^2+\frac{\W^2}{4\G^2}}.
\eeq
\end{widetext}
We see that majority of the contribution in the integral comes
from $\W\sim\w_q$. So one may cutoff the $\W$-integral by $V\ll\G$. If
we cutoff $q$ at $q_c\sim \frac{V}{c_0}$, $\W_c\sim V\ll\G$. 
The static expression for the Keldysh polarization function must
be made consistent with the approximate dynamic dielectric function. For $\W\le V$,
\beq
-Im\Pi^K(Q)\approx D_0\square{\frac{\G_L^2+\G_R^2}{\G^3}\abs{\W}+
\frac{2\G_L\G_R}{\G^3}V}.
\eeq
After a series of integrals, we get
\beq
-Im\S_{1}^R(V)=\frac{G^2D_0\g^{neq}_{1}}{\G^2c_0^4}\square{1+\frac{2\G_L\G_R}{\G^2}}V^4,
\eeq
where $\g^{neq}_{1}\sim O(1)$. This gives us a fourth order contribution and will
become a subdominant correction to the second integral as we will now show.
The second integral from Eq.\ref{se2} can be done straightforwardly:
\Beq
-Im\S^R_2(V)&\approx&\frac{1}{\G^2}\int_0^{V/c_0}\frac{qdq}{2\p}
\abs{g_{sc}({\bf q})}^2\G\nonumber\\
&\times&\round{sgn(\w_q-V/2)+sgn(\w_q+V/2)}\nonumber\\
&=&\frac{GV^3}{\G c_0^3}\underbrace{\int_{1/2}^{1}\frac{y^2dy}{2\p}}_{\g^{neq}_2}\nonumber\\
&=&\frac{G\g^{neq}_2}{\G c_0^3}V^3,
\Eeq
where $\g^{neq}_{2}\sim O(1)$. Therefore, to lowest order in voltage we obtain,
\beq
{1\over\ta_{el-ph}^{neq}}\approx\frac{G\g^{neq}}{\G c_0^3}V^3,
\label{srv}
\eeq
where $\g^{neq}\sim O(1)$.

We can summarize the low-$T$/low-$V$ electron-phonon scattering rate both in and 
out of equilibrium:
\Beq
{1\over\ta_{el-ph}}&=&\frac{G\g_X}{\G c_0^3}X^3\nonumber\\
&\mbox{where}&\left\{\begin{array} {r@{\quad :\quad}l}X=T & T\ne 0, V=0 \\ 
X=\abs{V} & T=0, V\ne 0\end{array}\right., 
\Eeq 
where $\g_X$ is some real constant of $O(1)$.

%%%%%%%%%%%%%%%%%%%%%%%%%%%%%%%%%%%%%%%%%%%%%%%%%%%%%%%%%%%%%%%%%%%%%%%%%%%%%%%%%%%%
\subsection{Electron Mass Enhancement}
We now compute the effective on-shell mass correction and investigate the 
effects of temperature and voltage to lowest order. 

%%%%%%%%%%%%%%%%%%%%%%%%%%%%%%%%%%%%%%%%%%%%%%%%%%%%%%%%%%%%%%%%%%%%%%%%%%%%%%%%%%%%%%%%
\subsubsection{Thermal Equilibrium $T\ne 0$, $V=0$}
From Eqs.\ref{sth1},\ref{sth2} we obtain,
\bml
Re\S_1^R(\w,T)\\=2\int\frac{qdq}{2\p}\abs{g_{sc}({\bf q})}^2
\coth\round{\frac{\w_q}{2T}}f_1(q)\frac{\w}{\G},
\label{resig1}
\end{multline}  
\bml
Re\S_2^R(\w,T)=2\int\frac{qdq}{2\p}\int\frac{d\W}{2\p}\abs{g_{sc}({\bf q})}^2\\
\times\tanh\round{\frac{\w+\W}{2T}}f_0(q)P\round{\frac{2\w_q}{\W^2-\w_q^2}}.
\label{resig2}
\end{multline}
It is useful to separate out the zero and finite temperature contributions to the mass
enhancement by defining
\beq
\la_i(T)=\la_i^0+\de\la_i(T),
\eeq
where
\beq
\la_i^0\equiv -\left.\pder{Re\S_i^R(\w,T=0)}{\w}\right|_{\w=0},
\eeq
\bml
\de\la_i(T)\equiv -\left[\left.\pder{Re\S_i^R(\w,T)}{\w}\right|_{\w=0}\right.\\
\left.-\left.\pder{Re\S_i^R(\w,T=0)}{\w}\right|_{\w=0}\right].
\end{multline}
Here, $i\in\{1,2\}$ labels the two self-energy contributions 
(Eqs.\ref{resig1},\ref{resig2}). The zero temperature contribution is then given by,
\bml
\la^0=\sum_{i=1}^2\la_i^0\\=\frac{2G}{\p^2}\int_0^{q_D\sim k_F}q^2dq\frac{f_0(q)}{\w_q}
\round{1-\frac{\p}{2}\frac{f_1(q)}{f_0(q)}\frac{\w_q}{\G}}.
\end{multline}
We see that the second term in the round brackets is much smaller than 1 for $q\le k_F$.
So we neglect this term and obtain the zero temperature enhanced mass in equilibrium:
\Beq
\la^0&\approx&\frac{2G}{\p^2c_0^2}\int_0^{k_F}qdqf_0(q)\nonumber\\
&=&\frac{2G\mi}{\p^2c_0v_F^2}.\qquad(\mi\gg\G)
\Eeq
At finite temperature, the effective mass correction gains a temperature correction. 
The first correction is given by,
\beq
\de\la_1(T)=-\frac{2G}{\G}\int\frac{q^2dq}{2\p}\square{\coth\round{\frac{\w_q}{2T}}-1}f_1(q).
\eeq
We see that the integrand is negligibly small for $q\ge\al\frac{T}{c_0}$ where $\al\sim O(1)$.
Then, employing this momentum cutoff we get,
\Beq
\de\la_1(T)&=&-\frac{2G}{\G}\int_0^{\al\frac{T}{c_0}}\frac{q^2dq}{2\p}\square{\coth\round
{\frac{c_0q}{2T}}-1}f_1(q)\nonumber\\
&\approx&-\frac{G}{\G^2}\round{\frac{T}{c_0}}^3\underbrace{\frac{1}{\p}\int_0^{\al}y^2dy
\square{\coth\round{\frac{y}{2}}-1}}_{l_3}\nonumber\\
&=&-\frac{G}{c_0^3\G^2}l_3 T^3,
\Eeq
where $l_3>0$ is a constant of $O(1)$. The second contribution is,
\bml
\de\la_2(T)=-2\int\frac{qdq}{2\p}\frac{d\W}{2\p}\abs{g_{sc}(q)}^2f_0(q)\\
\times\round{\frac{1}{2T}\mbox{sech}^2\round{\frac{\W}{2T}}-2\de(\W)}P
\round{\frac{2\w_q}{\W^2-\w_q^2}}.
\label{dl2}
\end{multline}
We see that the integrand is negligibly small for phonon frequencies $\W\ge\al T$ where
once again $\al\sim O(1)$. In the second term, this corresponds to a momentum cutoff
of $q_c=\al\frac{T}{c_0}$. Applying these cutoffs we obtain the second correction
\beq
\de\la_2(T)=\frac{G}{c_0^3\G}l_2T^2,
\eeq
where the constant $l_2\sim O(1)>0$ is given by
\bml
l_2=\frac{1}{2\p^2}\int_{-\al}^\al dx\mbox{sech}^2\round{\frac{x}{2}}\\
\times\round{\al^2+x^2\ln\abs{\frac{\al^2}{x^2}-1}}-\frac{\al^2}{\p^2}.
\end{multline}
In conclusion, the electron mass enhancement at thermal equilibrium to lowest order in 
temperature is
\beq
\la(T)=\frac{2G\mi}{\p^2c_0v_F^2}+\frac{G\G}{c_0^3}l_2\round{\frac{T}{\G}}^2.
\eeq
%%%%%%%%%%%%%%%%%%%%%%%%%%%%%%%%%%%%%%%%%%%%%%%%%%%%%%%%%%%%%%%%%%%%%%%%%%%%%%%%%%%%%
\subsubsection{Out of Equilibrium $T=0$, $V\ne 0$}
We begin the nonequilibrium calculation with the second contribution $\S_2^R(\w,V)$,
\bml
Re\S_2^R(\w,V)=2\sum_\al\frac{\G_\al}{\G}\int\frac{qdq}{2\p}\frac{d\W}{2\p}
\abs{g_{sc}({\bf q})}^2\\
\times sgn\round{\w+\W-\frac{V_\al}{2}}f_0(q)P\round{\frac{2\w_q}{\W^2-\w_q^2}}.
\end{multline}
Then the finite voltage correction is given by
\Beq
\de\la_2(V)&=&4G\int\frac{q^2dq}{2\p}\frac{d\W}{2\p}f_0(q)P\frac{2\w_q}{\W^2-\w_q^2}
\nonumber\\
&\times&\square{\de\round{\W}-\sum_\al\frac{\G_\al}{\G}\de\round{\frac{V_\al}{2}-\W}}
\nonumber\\
&=&\frac{G}{\G c_0^3}V^2\underbrace{\square{\frac{1}{\p^2}P\int_0^1dy\frac{4y}{4y^2-1}}}_{k_2}
\nonumber\\
&=&\frac{G\G}{c_0^3}k_2\round{\frac{V}{\G}}^2.
\Eeq
Here, $k_2\sim O(1)$ is a positive constant. For the calculation of $\de\la_1(V)$ one needs
to go beyond the static approximation for the dielectric function as was done for the
scattering rate calculation. Thus, the correction to the effective mass parameter from the 
first contribution is given by 
\Beq
\de\la_1(V)&=&-\frac{D_0}{\p^2\G}\frac{2\G_L\G_RV}{\G^3}\nonumber\\
&\times&\int qdqd\W f_1(q)\frac{\abs{g_q}^4\w_p^2}{\w_q^4}\frac{q^4}{k_S^4}\frac{1}
{\round{1-\frac{\W^2}{\w_q^2}}^2+\frac{\W^2}{4\G^2}}\nonumber\\
&\approx&-\frac{D_0}{\p^2\G}\frac{2\G_L\G_R}{\G^2}\round{\frac{V}{\G}}^2\int_0^{V/c_0} 
qdq\frac{G^2q^2}{c_0^2q^2}\nonumber\\
&=&-\frac{D_0G^2\G_L\G_R}{\p^2\G c_0^4}\round{\frac{V}{\G}}^4.
\Eeq
The electron mass enhancement to lowest order in voltage is then,
\beq
\la(V)=\frac{2G\mi}{\p^2c_0v_F^2}+\frac{G\G}{c_0^3}k_2\round{\frac{V}{\G}}^2.
\eeq
In conclusion, the electron mass enhancement to lowest order in $\frac{T}{\G}$ or 
$\frac{V}{\G}$, can be summarized concisely by
\Beq
\la(X)&=&\frac{2G\mi}{\p^2c_0v_F^2}+\frac{G\G}{c_0^3}\ka_X\round{\frac{X}{\G}}^2\nonumber\\
&\mbox{where}&\left\{\begin{array} {r@{\quad :\quad}l}X=T& T\ne 0, V=0 \\ 
X=V& T=0, V\ne 0\end{array}\right.,
\Eeq
where $\ka_X$ is a real constant of $O(1)$.

%%%%%%%%%%%%%%%%%%%%%%%%%%%%%%%%%%%%%%%%%%%%%%%%%%%%%%%%%%%%%%%%%%%%%%%%%%%%%%%%%%%%%%%%%
\section{Conclusion}
\label{conclude}
In this paper we studied a simple model of a steady-state electrically-driven two-dimensional 
electron-phonon system. The drive was applied to the metallic layer by attaching two 3D leads,
which acted both as a source of particles (current source) and a heat sink. The resultant
current was perpendicular to the layer so that the heating problem could be avoided. 
The effective theory for the metallic layer was developed using the Keldysh path integral method.

We found that various properties of the electron-phonon system is modified in the
presence of an out-of-plane voltage. Voltage dependences in the Thomas-Fermi screening length and 
in the velocity of the longitudinal acoustic phonon mode are presented. 
The results show that both of these quantities can be tuned at will using the
external voltage. In-plane electron-phonon scattering rate and electron mass enhancement were also 
investigated. We showed that electron-phonon scattering can be enhanced by voltage 
at zero temperature. The computed modification to the electron mass enhancement by voltage 
implies the possibility of tuning the effective mass of the electron using voltage.

The in-plane electron-phonon scattering rate can be indirectly measured by observing
the voltage-dependent piece of the in-plane resistivity, $\rho^V_{xx}$. The resistivity
can be measured as a linear response to the in-plane current drive. Since the electron-phonon
scattering rate occurs predominantly in the forward scattering channels, $\rho^V_{xx}$ gains an
additional factor of $V^2$ compared to the scattering rate. Therefore, the finite voltage
correction to the in-plane resistivity should scale as $\rho^V_{xx}\sim V^5$.

Renormalizations to the electron effective mass and electron distribution due to external voltage 
may be observed in a Shubnikov-de-Haas experiment.
In the presence of a magnetic field normal to the metallic layer (${\bf B}=B\hat{{\bf z}}$),
the layer electrons undergo an in-plane cyclotron motion. The electron energies are quantized 
according to an energy spectrum composed of Landau levels, 
which are separated by the cyclotron energy. The resultant electron density of
states show multiple peaks, each peak corresponding to a Landau level. Because the separation
between a pair of Landau levels depends linearly on magnetic field, the longitudinal in-plane 
resistivity oscillates as a function of the magnetic field. This oscillation is the well-known 
Shubnikov-de-Haas effect. The difference in the extrema of the oscillations as a function of magnetic 
field decays exponentially as the field is decreased\cite{coleridgeetal,mancoffetal}. 
The cyclotron frequency can be obtained by observing the characteristic energy scale with 
which this decay occurs. The effective electron mass, in turn, can be found by its direct 
relationship to the cyclotron frequency. 

We also predict that the observed Shubnikov-de-Haas oscillations should contain features that
specifically result from the structure of the zero-temperature nonequilibrium electron distribution.
Out of equilibrium, the distribution function assumes the split-step shape\cite{comm}, in which each step 
is presumably of different height given the asymmetry in the lead-layer couplings. As the magnetic
field is varied, both of these steps become resonant with a Landau level. As a result, each 
extremum in the resistivity would, in principle, split into two asymmetric peaks.

\acknowledgments
We would like to thank S. Julian for a helpful discussion.
This work was supported by the NSERC of Canada, the Canada Research Chair program, 
the Canadian Institute of Advanced Research, and KRF-2005-070-C00044.

%%%%%%%%%%%%%%%%%%APPENDIX%%%%%%%%%%%%%%%%%%%%%%%%%%%%%%%%%%%%%%%%%%%%%%%%%%%%
\newpage
\appendix
\begin{widetext}
\section{Keldysh Action for Phonons}
\label{details}
The purpose of this appendix is to begin with the real time action for the Hamiltonian
given in section \ref{model} and show detailed Keldysh calculations that lead to the 
effective phonon action Eq.\ref{su}, from which the inverse retarded, advanced and Keldysh phonon
propagators can be read off directly.

Our starting real time action, after the lead electrons have been integrated out, is
\beq
iS[d,\bar{d},u,u^*]=iS_{el}[d,\bar{d}]+iS_{ph}[u,u^*]+iS_{el-ph}[d,\bar{d},u,u^*],
\eeq
where
\beq\bsp
iS_{el}[d,\bar{d}]&=i\int dtdt'\sum_{{\bf k},\s}\bar{d}_\s({\bf k},t')G^{-1}({\bf
k},t'-t)d_\s({\bf k},t)\\
&-i\int dt \sum_{{\bf q}\ne 0}\sum_{{\bf k},{\bf k'}\atop \s,\s '}\frac{V^{ee}_{{\bf q}}}{2}
\bar{d}_\s({\bf k+q},t)\bar{d}_{\s'}({\bf k'-q},t)d_{\s '}({\bf k'},t)d_\s({\bf k},t),\\
iS_{ph}[u,u^*]&=\int dt u^*({\bf k},t')D_0^{-1}({\bf k},t'-t)u({\bf k},t),\\
iS_{el-ph}[d,\bar{d},u,u^*]&=-i\int dt\sum_{{\bf k},{\bf q},\s}g_{{\bf q}}\bar{d}_\s({\bf k+q},t)
d_\s({\bf k},t)u({\bf q},t).
\label{srealtime}
\end{split}\eeq
In order to carry out the time loop contour integration, we split all fields into two components,
forward ($+$) and backward ($-$) fields, which reside on the forward and backward branches of 
the time loop contour, respectively. The action then can be written in the form,
\beq
iS^K[d,\bar{d},u,u^*,\phi,\phi^*]=iS^K[d_+,\bar{d}_+,u_+,u^*_+,\phi_+,\phi^*_+]
-iS^K[d_-,\bar{d}_-,u_-,u^*_-,\phi_-,\phi^*_-].
\label{stimeloop}
\eeq
Here, $\phi$ is a Hubbard-Stratonovic field used to decouple the quartic term in Eq.\ref{srealtime}.
In this basis for the Keldysh space (i.e. $+$,$-$ fields) one obtains four Green functions for every
field, one of which can be expressed as a linear combination of the other three. Therefore, one
often performs a linear transformation of the fields so as to work with three independent Green
functions. This is known as Keldysh rotation\cite{kamenev,rammer}, in which new fermion
fields are defined via
\beq
f_1=\frac{f_++f_-}{2}\qquad f_2=\frac{f_+-f_-}{2}\qquad
\bar{f}_1=\frac{\bar{f}_+-\bar{f}_-}{2}\qquad\bar{f}_2=\frac{\bar{f}_++\bar{f}_-}{2},
\eeq
and new boson fields are defined via
\beq
b_{cl}=\frac{b_++b_-}{2}\qquad b_q=\frac{b_+-b_-}{2}\qquad
b_{cl}^*=\frac{b_+^*+b_-^*}{2}\qquad b_q^*=\frac{b_+^*-b_-^*}{2}.
\eeq
Upon carrying out the Keldysh rotation on the time loop contour action of Eq.\ref{stimeloop}
we obtain,
\beq\bsp
iS^{el}_K&=2i\int dt\sum_{{\bf k},{\bf k'},\s}\rvec{\bar{d}_\s^1({\bf k'},t)}
{\bar{d}_\s^2({\bf k'},t)}\left[\mat{G_R^{-1}({\bf k},t)\de_{{\bf k'}{\bf k}}}
{[\hat{G}^{-1}]_K({\bf k},t)\de_{{\bf k'}{\bf k}}}{0}{G_A^{-1}({\bf k},t)\de_{{\bf k'}{\bf k}}}
\right.\\
&+\left.\mat{e\phi^{cl}({\bf k'}-{\bf k},t)-g_{{\bf k'}-{\bf k}}u^{cl}({\bf k'}-{\bf k},t)}
{e\phi^{q}({\bf k'}-{\bf k},t)-g_{{\bf k'}-{\bf k}}u^{q}({\bf k'}-{\bf k},t)}
{e\phi^{q}({\bf k'}-{\bf k},t)-g_{{\bf k'}-{\bf k}}u^{q}({\bf k'}-{\bf k},t)}
{e\phi^{cl}({\bf k'}-{\bf k},t)-g_{{\bf k'}-{\bf k}}
u^{cl}({\bf k'}-{\bf k},t)}\right]\cvec{d_\s^1({\bf k},t)}{d_\s^2({\bf k},t)}\\
iS^{ch}_K&=2i\int dt\sum_{{\bf k}}\rvec{\phi^{cl*}({\bf k},t)}{\phi^{q*}({\bf k},t)}
\mat{0}{\frac{k}{2\p}}{\frac{k}{2\p}}{0}\cvec{\phi^{cl}({\bf k},t)}
{\phi^{q}({\bf k},t)}\\
iS^{ph}_K&=2i\int dt\sum_{{\bf k}}\rvec{u^{cl*}({\bf k},t)}{u^{q*}({\bf k},t)}
\mat{0}{[D_0^A({\bf k},t)]^{-1}}{\square{D_0^R({\bf k},t)}^{-1}}
{\square{\hat{D}_0^{-1}({\bf k},t)}^{K}}\cvec{u^{cl}({\bf k},t)}{u^{q}({\bf k},t)}.
\end{split}\eeq
Integrating out the layer electrons in $iS^{el}_K$, we get
\beq
iS^{eff}_K[\Phi,\Phi^*]=-2i\sum_{{\bf k},\w}\rvec{\Phi^{cl*}({\bf k},\w)}
{\Phi^{q*}({\bf k},\w)}\mat{0}{\Pi^A({\bf k},\w)}{\Pi^R({\bf k},\w)}
{\Pi^K({\bf k},\w)}\cvec{\Phi^{cl}({\bf k},\w)}{\Phi^{q}({\bf k},\w)}.
\label{skelPhi}
\eeq
When Eq.\ref{skelPhi} is expanded, the total action now becomes,
\beq\bsp
iS_k^{eff}[u,u^*,\phi,\phi^*]&=2i\sum_{{\bf k},\w}\rvec{u^{cl*}({\bf k},\w)}{u^{q*}({\bf k},\w)}
\mat{0}{[D_0^A({\bf k},\w)]^{-1}}{\square{D_0^R({\bf k},\w)}^{-1}}{\square{D_0^{-1}({\bf k},\w)}^{K}}
\cvec{u^{cl}({\bf k},\w)}{u^{q}({\bf k},\w)}\\
&+2i\sum_{{\bf k},\w}\rvec{\phi^{cl*}({\bf k},\w)}{\phi^{q*}({\bf k},\w)}
\mat{0}{\frac{k}{2\p}}{\frac{k}{2\p}}{0}\cvec{\phi^{cl}({\bf k},\w)}{\phi^{q}({\bf k},\w)}\\
&-2ie^2\sum_{{\bf k},\w}\rvec{\phi^{cl*}({\bf k},\w)}{\phi^{q*}({\bf k},\w)}
\mat{0}{\Pi^A({\bf k},\w)}{\Pi^R({\bf k},\w)}{\Pi^K({\bf k},\w)}
\cvec{\phi^{cl}({\bf k},\w)}{\phi^{q}({\bf k},\w)}\\
&+2ie\sum_{{\bf k},\w}g^*_{{\bf k}}
\rvec{u^{cl*}({\bf k},\w)}{u^{q*}({\bf k},\w)}
\mat{0}{\Pi^A({\bf k},\w)}{\Pi^R({\bf k},\w)}{\Pi^K({\bf k},\w)}
\cvec{\phi^{cl}({\bf k},\w)}{\phi^{q}({\bf k},\w)}\\
&+2ie\sum_{{\bf k},\w}g_{{\bf k}}
\rvec{\phi^{cl*}({\bf k},\w)}{\phi^{q*}({\bf k},\w)}
\mat{0}{\Pi^A({\bf k},\w)}{\Pi^R({\bf k},\w)}{\Pi^K({\bf k},\w)}
\cvec{u^{cl}({\bf k},\w)}{u^{q}({\bf k},\w)}\\
&-2i\sum_{{\bf k},\w}|g_{{\bf k}}|^2
\rvec{u^{cl*}({\bf k},\w)}{u^{q*}({\bf k},\w)}
\mat{0}{\Pi^A({\bf k},\w)}{\Pi^R({\bf k},\w)}{\Pi^K({\bf k},\w)}
\cvec{u^{cl}({\bf k},\w)}{u^{q}({\bf k},\w)}.
\label{sphiu}
\end{split}\eeq
The $\Phi$ fields were defined in Eq.\ref{Phi} and the Keldysh polarization diagrams
($\Pi$) were defined in Eqs.\ref{pir},\ref{pik}. Expressions for the retarded and
Keldysh polarization diagrams in the long-time, long-wavelength limit were introduced in
Eqs.\ref{pir2},\ref{pik2}. The coefficients appearing in these approximate expressions 
are given by,
\Beq
c_1(V)&\equiv& \sum_{\al}\frac{\G_\al}{\G}D(\mi_\al)\\
c_2(V)&\equiv&\frac{D_0}{3\p}\sum_\al\frac{\G_\al}{\mi}\frac{\mi^2}{\mi_\al^2+\G^2}\\
c_3(V)&\equiv& \sum_\al\frac{\G_\al}{2\G}\square{D(\mu_\al)+\frac{D_0}{\p}\frac{\G}{\mu_\al}\frac{1}
{1+\frac{\G^2}{\mu_\al^2}}}\\
c_4&\equiv&D(\mi)+\frac{D_0\G}{\p\mi}\frac{1}{1+\frac{\G^2}{\mi^2}}.
\Eeq
Now integrating out the $\phi$ field, we obtain
\beq\bsp
iS_K^{eff}[u,u^*]&=2i\sum_{{\bf k},\w}\rvec{u^{cl*}({\bf k},\w)}{u^{q*}({\bf k},\w)}
\mat{0}{[D_0^A({\bf k},\w)]^{-1}}{\square{D_0^R({\bf k},\w)}^{-1}}
{\square{D_0^{-1}({\bf k},\w)}^{K}}
\cvec{u^{cl}({\bf k},\w)}{u^{q}({\bf k},\w)}\\
&-2ie^2\sum_{{\bf k},\w}|g_{{\bf k}}|^2\rvec{u^{cl*}({\bf k},\w)}{u^{q*}({\bf k},\w)}
\mat{0}{\Pi^A({\bf k},\w)}{\Pi^R({\bf k},\w)}{\Pi^K({\bf k},\w)}\\
&\times\mat{0}{\frac{k}{2\p}-e^2\Pi^A({\bf k},\w)}
{\frac{k}{2\p}-e^2\Pi^R({\bf k},\w)}{-e^2\Pi^K({\bf k},\w)}^{-1}
\mat{0}{\Pi^A({\bf k},\w)}{\Pi^R({\bf k},\w)}{\Pi^K({\bf k},\w)}
\cvec{u^{cl}({\bf k},\w)}{u^{q}({\bf k},\w)}\\
&-2i\sum_{{\bf k},\w}|g_{{\bf k}}|^2\rvec{u^{cl*}({\bf k},\w)}{u^{q*}({\bf k},\w)}
\mat{0}{\Pi^A({\bf k},\w)}{\Pi^R({\bf k},\w)}{\Pi^K({\bf k},\w)}
\cvec{u^{cl}({\bf k},\w)}{u^{q}({\bf k},\w)}\\
&=2i\sum_{{\bf k},\w}\rvec{u^{cl*}({\bf k},\w)}{u^{q*}({\bf k},\w)}\left[
\mat{0}{[D_0^A({\bf k},\w)]^{-1}}{\square{D_0^R({\bf k},\w)}^{-1}}
{\square{D_0^{-1}({\bf k},\w)}^{K}}\right.\\
&-\left.e^2|g_{{\bf k}}|^2\mat{0}{\frac{[\Pi^A({\bf k},\w)]^2}
{\frac{k}{2\p}-e^2\Pi^A({\bf k},\w)}}{\frac{[\Pi^R({\bf k},\w)]^2}
{\frac{k}{2\p}-e^2\Pi^R({\bf k},\w)}}{\frac{\Pi^K({\bf k},\w)}
{e^2}\round{\frac{1}{|\e({\bf k},\w)|^2}-1}}-|g_{{\bf k}}|^2
\mat{0}{\Pi^A({\bf k},\w)}{\Pi^R({\bf k},\w)}{\Pi^K({\bf k},\w)}\right]
\cvec{u^{cl}({\bf k},\w)}{u^{q}({\bf k},\w)}.
\label{isph}
\end{split}\eeq
We see that our effective phonon Keldysh action obeys the causality structure for bosons. The 
inverse retarded, advanced, and Keldysh propagators can now be read off directly from this
action. The results were presented in Eqs.\ref{dr},\ref{dk}.
\vspace{3cm}
\end{widetext}

%%%%%%%%%%%%%%%%%%%%%%%%%%%%%%%%%%%%%%%%%%%%%%%%%%%%%%%%%%%%%%%%%%%%%%%%%%%%%%%%%%%%%%%%%%%%


\begin{thebibliography}{99}
\bibitem{dalphil} D. Dalidovich and P. Phillips, Phys. Rev. Lett. {\bf 93}, 27004 (2004).
\bibitem{feldman} D. Feldman, Phys. Rev. Lett. {\bf 95}, 177201 (2005).
\bibitem{greensondhi} A.G. Green and S.L. Sondhi, Phys. Rev. Lett. {\bf 95}, 267001 (2005).
\bibitem{greenetal} A.G. Green, J.E. Moore, S.L. Sondhi, and A.Vishwanath, Phys. Rev. Lett.
{\bf 97}, 227003 (2006).
\bibitem{hogangreen} P.M. Hogan and A.G. Green, condmat/0607522.
\bibitem{calleg} A.O. Caldeira and A.J. Leggett, Ann. Phys. (NY) {\bf 149}, 374 (1983).
\bibitem{keldysh} L.V. Keldysh, Zh. Eksp. Teor. Fiz. {\bf 47}, 1515 (1964). 
[Sov. Phys. JETP {\bf 20}, 1018 (1965).]
\bibitem{rammer} J. Rammer and H. Smith, Rev. Mod. Phys. {\bf 58}, 323 (1986).
\bibitem{kamenev} A. Kamenev, condmat/0412296.
\bibitem{rev} L.P. Kouwenhoven, C.M. Marcus, P.L. McEuen, S. Tarucha, R.M. Westervelt, and
N.S. Wingreen, {\it Proceedings of the NATO Advanced Study Institute on Mesoscopic Electron Transport},
ed. L.L. Sohn, L.P. Kouwenhoven, and G. Sch\"{o}n, Kluwer Series E345 pp.105-214 (1997).
\bibitem{jauhoetal} A. Jauho, N.S. Wingreen, and Y. Meir, Phys. Rev. B {\bf 50}, 5528 (1994).
\bibitem{rosch} A. Rosch, J. Kroha, and P. W\"{o}lfle, Phys. Rev. Lett. {\bf 87}, 156802 (2001)
\bibitem{mitra1} A. Mitra, I. Aleiner, A.J. Millis, Phys. Rev. B {\bf 69}, 245302 (2004).
\bibitem{mitra2} A. Mitra, I. Aleiner, A.J. Millis, Phys. Rev. Lett. {\bf 94}, 76404 (2005).
\bibitem{mitra3} A. Mitra, S. Takei, Y.B. Kim, A.J. Millis, Phys. Rev. Lett.
                 {\bf 97}, 236808 (2006).
\bibitem{little2} M.H. Szyma\'{n}ska, J. Keeling, and P.B. Littlewood, Phys. Rev. Lett. 
{\bf 96}, 230602 (2006).
\bibitem{little} A.H. Szyma\'{n}ska, J. Keeling, and P.B. Littlewood, condmat/0611456.
\bibitem{bohmstaver} D. Bohm and I. Staver, Phys. Rev. {\bf 84}, 836 (1951).
\bibitem{coleridgeetal} P.T. Coleridge, R. Stoner, and R. Fletcher, Phys. Rev. B {\bf 39}, 
1120 (1989).
\bibitem{mancoffetal} F.B. Mancoff, L.J. Zielinski, C.M. Marcus, K. Campman, and A.C. Gossard,
Phys. Rev. B {\bf 53}, R7599 (1996).
\bibitem{comm} As mentioned earlier, the two steps in the nonequilibrium split-step 
electron distribution function are broadened in the presence of electron-phonon interaction. 
However, in the small voltage regime, where our theory is valid, the two steps may still 
be fairly well-defined since the electron-phonon broadening effects are of order $V^3$, 
and are small compared to the split size of the two steps, which is of order $V$. 
\end{thebibliography}
\end{document}